# Beyond the Interface Limit: Structural and Magnetic Depth Profiles of Voltage-Controlled Magneto-Ionic Heterostructures

Dustin A. Gilbert,[1*] Alexander J. Grutter,[1*] Elke Arenholz,[2] Kai Liu,[3] B. J. Kirby,[1]

Julie A. Borchers,[1] Brian B. Maranville[1]

[1]*NIST Center for Neutron Research, National Institute of Standards and Technology, Gaithersburg, MD 20899*

[2]*Advanced Light Source, Lawrence Berkeley National Laboratory, Berkeley, CA 94720, USA*

[3]*Physics Department, University of California, Davis, CA 95616, USA*

[*]These authors contributed equally to this work

**Abstract**

Electric-field control of magnetism provides a promising route towards ultralow power information storage and sensor technologies. The effects of magneto-ionic motion have so far been prominently featured in the direct modification of interface chemical and physical characteristics. Here we demonstrate magnetoelectric coupling moderated by voltage-driven oxygen migration beyond the interface limit in relatively thick $AlO_x/GdO_x/Co$ (15 nm) films. Oxygen migration and its ramifications on the Co magnetization are quantitatively mapped with polarized neutron reflectometry under thermal and electro-thermal conditionings. The depth-resolved profiles uniquely identify interfacial and bulk behaviors and a semi-reversible suppression and recovery of the magnetization. Magnetometry measurements show that the conditioning changes the microstructure so as to disrupt long-range ferromagnetic ordering, resulting in an additional magnetically soft phase. X-ray spectroscopy confirms electric field induced changes in the Co oxidation state but not in the Gd, suggesting that the $GdO_x$ transmits oxygen but does not source or sink it. These results together provide crucial insight into controlling magnetic heterostructures via magneto-ionic motion, not only at the interface, but also throughout the bulk of the films.

**Introduction**

With exciting potentials of energy-efficiency and new functionalities in hybrid magnetoelectric devices, voltage control of magnetism is currently the focus of intense investigations.[1-5] Traditional spintronic architectures use electron spin for information storage and transmission, e.g., utilizing the giant[6,7] or tunneling[8,9] magnetoresistance, spin transfer torque,[10-12] or the spin Hall effects.[13,14] Recently, an alternative magneto-ionic approach has been demonstrated, e.g., in $GdO_x$/Co Hall bar structures, that uses an electro-thermal conditioning sequence to drive oxygen from an oxide film into a neighboring ferromagnetic metal layer[15-18] and directly tailor the interface chemistry. In so doing, *interfacial* physical properties such as magnetic anisotropy and saturation magnetization are effectively controlled by the electric field, even *reversibly*. The most prominent effect of the magneto-ionic motion is the modification of the interface - indeed the ultra-thin Co films in these studies are less than five monolayers thick, and are consequently dominated by the interfacial properties. Considering the strong reduction potential of gadolinium[19] and relatively good chemical stability of cobalt oxide[20], the observed reversibility cannot be explained by simple enthalpy considerations alone, suggesting that the bulk and interfacial behaviors may be different. However, probing the magneto-ionic motion in the bulk of the film is intrinsically challenging.

In this work we report direct mapping of structural and magnetic depth profiles in voltage-controlled magneto-ionic heterostructures with relatively thick (15 nm) films of Co, and demonstrate that the electrically-induced oxygen migration extends far beyond the interface. Using polarized neutron reflectometry (PNR) the electrically-induced oxygen migration, manifest as increased nuclear scattering length density (SLD) and reduced magnetic SLD, is directly mapped. PNR depth profiles show that the electric field drives oxygen deep into the Co film (>10 nm), reducing the magnetization by more than 80% at the interface and 38% in the bulk. After reversing the polarity of the applied electric field, the magnetization recovers to 92% of the original value throughout the Co layer. Thus interface and 'bulk' behaviors are demonstrated to be different. Using the first order reversal curve (FORC) technique, we identify two distinct magnetic phases in the post-conditioned sample. This behavior differs significantly from that of the as-grown state, which exhibits a single magnetic phase, suggesting that the oxygen migration significantly alters the magnetic characteristics by impeding inter-granular coupling. Control experiments performed in the absence of an electric field reveal that annealing alone causes much smaller structural and magnetic changes to the Co film, distinguishing the roles of thermal conditioning from electro-thermal ones. X-ray absorption (XA) spectroscopy and X-ray magnetic circular dichroism (XMCD) confirm the role of the electric field in recovering the magnetization and cobalt oxidation states. XA performed on the $GdO_x$ shows that the Gd oxidation state is invariant, suggesting that the Gd transmits the oxygen rather than surrendering it.

**Results**

*Polarized Neutron Reflectometry*

Thin film samples of $Si/Pd(50~nm)/Al_2O_3(1~\mu m)/GdO_x(2~nm)/Co(15~nm)/Pd(20~nm)$ were grown by sputtering and evaporation. The samples were electro-thermally conditioned by sequentially applying +40 V then -40 V across the Pd layers at 230° C for 15 min (see Methods). PNR measurements of the as-grown sample and each conditioned state are shown in Fig. 1(a). The $R^{++}$ and $R^{--}$ reflectivities show sensitivity to the nuclear and magnetic depth profiles, evident by spin-dependent oscillations. The difference in the $R^{++}$ and $R^{--}$ is approximately proportional to the quotient of the magnetization and the nuclear SLD (see Methods). Thus, the magnetic contribution to the data is highlighted by plotting the spin asymmetry [SA = ($R^{++}$- $R^{--}$)/($R^{++}$+ $R^{--}$)], as shown in Fig. 1(b). The oscillation amplitude first decreases after conditioning in +40 V then increases after conditioning in -40 V, suggesting a decrease of the saturation magnetization, $M_S$, and/or a change in the structure, followed by a partial recovery towards the initial state. The nuclear and magnetic depth profiles from the converged model, shown in Fig. 1(c), confirm these trends.

The extracted depth profile of the as-grown sample accurately reproduces the designed structure, both in terms of thickness and nuclear SLD, $\rho_N$. Our fits show excellent agreement between the measured and expected $\rho_N$ values of Co ($2.27\times10^{-4}$ nm$^{-2}$), Pd ($4.02\times10^{-4}$ nm$^{-2}$), and GdO$_x$ ($2.74\times10^{-4}$ nm$^{-2}$)[21, 22]. However, the measured SLD of the thick Al$_2$O$_3$ base layer is substantially lower than the expected bulk value ($5.67\times10^{-4}$ nm$^{-2}$), suggesting the presence of significant voids or an oxygen deficient stoichiometry. The GdO$_x$ - a neutron absorber - can be identified explicitly by the imaginary SLD in Fig. 1(c). After conditioning the sample at +40 V, the nuclear SLD of the Co layer, $\rho_N^{Co}$, increases by 34%, approaching that of CoO ($4.29\times10^{-4}$ nm$^{-2}$), which is consistent with incorporation of oxygen. Simultaneously, the GdO$_x$/Co interface becomes much broader, increasing from a width of 3.3 nm to 10 nm and extending well into the Al$_2$O$_3$. After switching the voltage polarity to -40 V, the GdO$_x$/Co interface width is reduced to 1.9 nm and $\rho_N^{Co}$ decreases, demonstrating an induced migration of oxygen from the CoO$_x$ into and through the GdO$_x$. The recovery of $\rho_N^{Co}$ occurs predominantly within the 10 nm nearest the GdO$_x$ interface, while the top 5 nm, near the Co/Pd interface, remains unchanged from the +40 V conditioned state. Thus, we observe oxygen ion migration throughout the thickness of the Co layer, but reversibly only within the 10 nm closest to the GdO$_x$/Co interface. However, at the GdO$_x$ interface after this second conditioning, $\rho_N^{Co}$ is still 32% larger than the as-grown sample, demonstrating the oxygen migration is only semi-reversible.

Trends in the magnetic depth profile [dashed lines in Fig. 1(c)] agree with those observed in the nuclear profile. Specifically, the as-grown sample has a sharp step-function like GdO$_x$/Co (magnetic) interface and strong magnetic scattering, as indicated by its large magnetic SLD, $\rho_M$, of 3.45×10$^{-4}$ nm$^{-2}$ in the Co layer, corresponding to a bulk magnetization of 1180 emu cm$^{-3}$ (1 emu cm$^{-3}$ = 1 kA m$^{-1}$). Strikingly, the shape of the magnetic profile changes significantly after conditioning in +40 V, with $\rho_M$ reduced dramatically - by 80% - at the GdO$_x$/Co interface and 38% in the 'bulk'. Subsequent conditioning at -40 V recovers the original shape and $\rho_M$ increases to 92% of the as-grown value.

Control experiments were performed on a sample grown side-by-side with the electro-thermally conditioned sample, following the same thermal treatment but *without an electric field*. PNR measurements of the thermal-only sample, Fig. 2, show that the first thermal treatment also increases $\rho_N$ and reduces $\rho_M$ in the Co layer, but to a much lesser degree and with no significant changes in the GdO$_x$/Co interface shape and extent. Quantitative comparison shows that $\rho_N^{Co}$ increases by 17% and $\rho_M^{Co}$ decreases by 12% after thermal treatment alone compared to the as-grown sample. This is much less than the electro-thermally treated sample, which showed an increase in $\rho_N^{Co}$ of 34% and a decrease in $\rho_M^{Co}$ of 38% after the +40 V treatment. After a second 15 min thermal conditioning the nuclear and magnetic profiles of the control sample do not change appreciably, suggesting a saturation effect or depletion of easily diffusible oxygen. These results confirm the role of the electric field in enhancing the oxidation of the Co layer during the +40 V conditioning and reducing it during the -40 V treatment.

*Magnetometry*

Magnetic hysteresis loops of the samples as-grown, after both +/- 40 V treatment (E+T), and after two thermal-only treatments are shown in Fig. 3(a). The Co $M_S$ in the as-grown sample was measured to be 1230 emu cm$^{-3}$, in good agreement with the PNR value of 1180 emu cm$^{-3}$. Further, $M_S$ is decreased by 10% in the E+T treated sample and 7% in the thermal-only sample compared to the as-grown sample, similar to the PNR data which show a reduction of 10%. The good agreement between the magnetometry and PNR results supports the validity of the model used to fit the data. Sample coercivity and remanent magnetization are also decreased compared to the as-grown sample by 68% and 55% in the E+T sample and 54% and 11% in the thermal sample, respectively, indicating significant changes in the magnetic characteristics.

Details of the magnetization reversal have been investigated by the FORC method. The FORC distribution and family of FORCs for the as-grown sample are shown in Fig. 3(b) and inset, respectively. The family of FORCs show the minor loops fill the major loop, and the calculated FORC distribution shows only a single feature, centered at (local coercivity $\mu_0 H_C$ = 4.9 mT, bias field $\mu_0 H_B$ = 0 mT), indicating the sample is comprised of a single magnetic phase.[23, 24] After the combined +/- 40 V electric field treatment (E+T) the family of FORCs still fill the major loop, but the FORC distribution, Fig. 3(c), now shows two features. The main feature is centered at ($\mu_0 H_C$ = 2.6 mT, $\mu_0 H_B$ = 0.33 mT) and is circularly-symmetric, again indicative of an irreversible (i.e. hysteretic) phase. The shift in $H_C$ relative to the as-grown sample indicates the coercivity is significantly reduced. The non-zero bias suggests a finite interaction experienced by this phase and may be the result of exchange bias with residual antiferromagnetic CoO with an enhanced Neel temperature.[25] A second phase is identified by the elongated FORC ridge along the $\mu_0 H_B$ axis centered at $\mu_0 H_C$ = 0 mT. This ridge represents a reversible phase with an internal demagnetizing interaction of 6 mT, identified by the spread of the feature along the $H_B$ axis.[26] A negative feature centered at ($\mu_0 H_C$ = 0.8 mT, $\mu_0 H_B$ = -1.7 mT) identifies reversal events which are present on FORC branches that start at $H_R^1$, $M(H,H_R^1)$, but absent on FORCs that begin at more negative $H_R^2$, $M(H,H_R^2)$ with $H_R^2 < H_R^1$.[23] In this case, the negative feature in Fig. 3(c) aligns in $H_R$ with the peak of the reversible feature, and in H with the irreversible feature. This behavior indicates that once the irreversible switching event occurs, the reversible phase changes its up-switching field due to magnetic coupling between the reversible and irreversible phases.

The family of FORCs for the thermally treated sample, Fig. 3(d) insert, is significantly different, with the minor loop protruding outside of the major loop. This indicates that the domain structure evolves more-easily under fields applied along the major loop that increase from the saturated state, than under fields that increase from the mixed multi-domain state.[27] This result further underscores the role of the electric field in determining the oxygen distribution. Similar to the E+T sample, the FORC distribution for the thermal-only sample, Fig. 3(d), also exhibits reversible and irreversible phases. The main FORC feature is centered at ($\mu_0 H_C$ = 2.6 mT, $\mu_0 H_B$ = 0.17 mT) and is no-longer circularly symmetric, but rather has a 90° bend with symmetries along the +H and -$H_R$ axes, typical of a domain nucleation/growth reversal mechanism.[28, 29] The FORC distribution for the thermal-only sample shows the same negative feature that again identifies magnetic coupling, and a new feature associated with the observed major loop protrusion. Integrating the FORC features gives a magnetic phase fraction:[24] the reversible phase contributes to 0%, 31%, and 24% of the magnetization in the as-grown, E+T, and thermal-only samples, respectively.

Reversible phases exhibit no hysteresis, and therefore are manifested in the FORC

distribution along the $\mu_oH_C = 0$ axis, e.g., when the phase has essentially zero coercivity or when the magnetic field is applied along the magnetic hard axis. Major hysteresis loops measured in the out-of-plane direction (see Supplemental Material) for these samples display little hysteresis, implying that the out-of-plane direction remains the hard axis. These results suggest that oxygen migrates in the film after E+T or thermal-only conditioning and segregates to grain boundaries in the Co layer, thus disrupting long-range magnetic correlations and effectively breaking down the affected Co films into isolated grains. Once the coupling between the grains becomes weaker, their respective magnetocrystalline anisotropies in confined grains play a large role in determining the magnetic orientation resulting in much reduced coercivity and remanent magnetization.

*X-ray Absorption and Circular Dichroism*

Oxidation of the cobalt after both +/- 40 V electro-thermal (E+T) and thermal-only treatments is confirmed in the XA and XMCD measurements (see Methods) shown in Fig. 4(a). Oxidation of the Co layer[30] is identified directly by the emergence of peaks at E = 779.2 eV and 776.8 eV, which are not present in the as-grown profile. The peak at 779.2 eV is largest in the thermal-only sample, indicating significantly increased oxidation relative to the E+T sample. This trend is supported by the XMCD spectra, which shows that the as-grown sample has the largest dichroism, indicating the largest magnetization. The E+T sample has the second largest dichroism, and the thermal-only sample has the smallest. The different ordering in the dichroism compared to the bulk magnetometry may identify variation in the depth-dependent oxygen binding behavior. This is consistent with the XA results, which showed a larger oxidation peak in the thermal-only treatment than the E+T sample. XMCD signal from the Gd, shown in Fig. 4(b), shows no dichroism for all three samples, indicating a negligible contribution to the magnetization. Interestingly, the XA signal for the Gd shows no significant change for any of the samples, suggesting a relatively constant Gd oxidation state, regardless of oxygen migration into or out of the Co.

**Discussion**

The PNR, magnetometry and X-ray results clearly demonstrate that electro-thermal conditioning can drive oxygen semi-reversibly into a thick (15 nm) Co film, profoundly changing its magnetic properties. Depth profiling with PNR indicates that while these effects are most prominent at the GdO/Co interface, they also extend throughout the entire 15 nm thick Co film. Reversing the polarity of the applied voltage drives oxygen out of the Co, partly restoring $\rho_N^{Co}$ and $\rho_M^{Co}$ to their original values at the GdO$_x$/Co interface, but leaving $\rho_N^{Co}$ and $\rho_M^{Co}$ unchanged near the Co/Pd interface. Thermal conditioning of the control sample also promotes oxidation of the Co layer, but the supply of highly mobile oxygen that can be moved by entropy-driven diffusion is clearly limited. In the following discussion we determine the oxygen stoichiometry from the nuclear scattering profile and consider the underlying mechanics of the oxygen migration.

*Oxygen Depth Profile*

The role of the electric field and entropy-driven oxygen migration is seen qualitatively by comparing the profiles for the +40 V electro-thermally treated sample with the thermally treated sample, Figs. 1(c) and 2(c) respectively. The thermal treatment is shown to scale the magnetic depth profile relative to the as-grown sample, but not change its shape. In comparison, the magnetic depth profile for the +40 V sample strongly deviates from that in the as-grown sample, suggesting that the electric field drives oxygen into the film, while the thermally activated, entropy-driven oxygen migration is relatively uniformly distributed.

Using the neutron coherent scattering length, $b$, for cobalt (2.49 fm) and oxygen (5.81 fm),[31] the $CoO_x$ stoichiometry can be directly calculated. Specifically, the nuclear SLD is calculated as: $SLD = (\frac{N_{Co}}{V} \cdot b_{Co} + \frac{N_O}{V} \cdot b_O)$ where $N_{Co(O)}$ is the total number of cobalt (oxygen) atoms within the volume, V, of the Co film. Assuming the as-grown film is pristine Co, which is supported by the good agreement with the referenced bulk $\rho_N^{Co}$ value, and that the cobalt number density remains constant during treatment, a lower-limit to the oxygen profile can be calculated: $\frac{N_O}{V} = \frac{SLD_{Measured} - SLD_{As-Grown}}{b_O}$. The stoichiometry is then defined as $CoO_{N(O)/N(Co)}$. The nuclear depth profiles suggest that the +40 V conditioning forms $CoO_{0.18}$, compared to $CoO_{0.07}$ in the thermally-treated sample. Treatment with a reversed polarity removes about one-third of the absorbed oxygen, leaving $CoO_{0.12}$. The fact that the magnetization measurements show only a 10% reduction in $M_S$ illustrates an indirect correspondence (e.g. not one-to-one) of the magnetization and nuclear composition and suggests some of the $O^{2-}$ may be forming magnetic compositions other than CoO, such as $Co_2O$, or may remain as interstitial oxygen. Integrating the oxygen profiles for the conditioned samples shows conservation of oxygen within this system to within 3%. Since we do not expect any external sources of oxygen, this agreement is another good support for the validity of the presented models. With additional assumptions we can further determine a depth-resolved oxygen profile (see Supplemental) to highlight the interface and bulk migration explicitly.

Interestingly, the depth profiles (especially the magnetic profile) indicate that the oxygen is semi-reversibly driven out of the $GdO_x$/Co (0-10 nm) interface after treatment in -40 V, while the remaining oxygen is left trapped deeper within the Co. We suggest that as the $GdO_x$/Co interface becomes depleted of oxygen, becoming more metallic. The oxidized region deeper within the film gets surrounded by conductive layers above and below, screening the electric field; without an electric field the oxygen does not migrate, resulting in the observed trapping effect. Thus, we suggest that the observed 10 nm thickness of the Co presents a practical limit for the electrically-driven oxygen migration within these otherwise metallic films.

*Mechanics of Oxygen Migration*

Brief considerations of the underlying mechanisms quickly reveal that this effect cannot be justified with bulk chemistry alone. First considering the thermally treated sample, the initial treatment both suppresses the magnetism and increases the nuclear SLD, consistent with inclusion of oxygen. Further treatment only weakly changed these parameters, indicating that all of the easily diffusible oxygen has migrated during the first treatment. Since the available thermal energy ($k_BT$ = 43 meV) is not enough to reduce $Gd_2O_3$ (enthalpy of formation, $\Delta Q$ = 16.1 eV), $Al_2O_3$ ($\Delta Q$ = 15.1 eV) or $CoO$ ($\Delta Q$ = 2.5 eV), we suggest the source of the mobile oxygen is likely interstitial, e.g., from trapped sputtering gas located at grain boundaries or voids in the $GdO_x$ and $Al_2O_3$, which diffuses and reacts with the Co (Co+O→CoO, $\Delta Q$ = -2.5 eV or $2Co+O_2$ → $2CoO$ $\Delta Q$ = -197 meV).[32, 33]

Next we consider the role of the electric field on oxygen mobility. Defect sites in oxygen-rich transition metal oxide films have been previously shown to act as p-type dopants,[34] giving them an effective negative charge. In the presence of an electric field with the anode on the Pd cap surface and cathode on the buried Pd seed film, excess $O^{2-}$ defects in the $GdO_x$ and $Al_2O_3$ will be pulled towards the cobalt, while vacancies are pulled towards the bottom Pd electrode. The chemical potential within each of the $Al_2O_3$, $GdO_x$ and Co films is expected to be uniform, and thus $O^{2-}$ defects are expected to be highly mobile within each.[17] The application of a static electric field then causes field-induced ion migration.[35] However, at the boundary between two layers there will exist a difference in the chemical potential, which may cause an accumulation of oxygen at e.g. the GdO$_x$/Co interface. Considering this issue, the enthalpy of formation is calculated in each layer and compared to the electric potential energy available to overcome this interfacial barrier. Assuming a lattice-site hopping model,[36, 37] the electric potential energy at the interface can be calculated to be 24 meV (qEΔx, where q is the oxygen charge of 2e$^-$, Δx is the atomic site spacing of 3 Å, and E is the electric field of 400 kV cm$^{-1}$). The sum of the electric potential energy and thermal energy defines the scale of the energy landscape available to drive the *reversible* oxygen migration back and forth across the interface. First considerations of an 'ideal' system were that the migrating oxygen is moved from chemically stable Gd$_2$O$_3$ to the Co (Gd$_2$O$_3$+3Co→3CoO+2Gd, ΔQ = +8.6 eV).[32, 33] Clearly the available energies (67 meV) cannot drive this reaction; stability of the Gd$_2$O$_3$ oxidation state is supported by the XA in Fig. 4. An alternative scenario suggested above is that the trapped oxygen react with the Co to form CoO (2Co+O$_2$ → 2CoO ΔQ = -197 meV). However, without a charge, the O$_2$ experiences no net force from the electric field, suggesting this reaction is not responsible for the observed effects induced by the electric field. Similar reactions with elemental oxygen (Co+O→CoO, ΔQ = -2.5 eV) can occur along the forward reaction, but will not be reversible as it again costs too much energy. Other considered reactions are presented in the supplemental materials, but no bulk energy calculation was able to support the observed results.

We present an alternative consideration (i.e. treating the grains as nanoscale clusters) that provides a reasonable energy landscape for the observed migration. In this picture, illustrated in Fig. 5, oxygen ions are bound to the surface of the nanocrystalline grains. These O$^{2-}$ ions are off-stoichiometry defects, and have a different binding energy than the core of the grain. Since the grains were all fabricated at the same time by sputtering in an oxygen-rich environment, the surface structure, and hence energy landscape, is expected to be relatively uniform. Thus applying an electric field moves the surface $O^{2-}$ towards the anode and into the Co-layer, forming $CoO_x$. Once on the grain surfaces these $O^{2-}$ ions experience chemically-driven diffusion, which occurs at a rate of approximately 10-20 nm/hr at these temperatures,[38] preferentially oxidizing the surface, but likely also the core of the small cluster-like Co grains. Reversing the applied field drives $O^{2-}$ to leave the $CoO_x$ film. Using a cluster-like formation approach, the binding energy per atom for $GdO_x$[39] and $CoO_x$[40] can be determined by modeling to be 4.50 eV/atom and 4.48 eV/atom, respectively, near nominal stoichiometry. Since the energy landscape is isotropic to within 20 meV, one only has to overcome the small barrier and the activation energy to move oxygen ions, destroy old bonds and create new ones. Thus oxygen freely leaves the $CoO_x$, but due to the slow bulk diffusion in the $CoO_x$ grains, a gradient oxygen profile is expected to form within the grains, with the surface being oxygen deficient. Once the grain surface loses enough oxygen to become conductive, the internal electric field is again screened, trapping the oxygen within the core. Thus, both at the film-length mesoscale and grain-length nanoscale, the oxygen migration induces screening effects which limits the oxygen mobility. We propose that, as a result, the conditioned films have a highly defective structure, with a complex mix of $CoO_x$ core-shell-surface grains, which is consistent with the two-phase construction identified by the FORC measurement. Reconciling this with the XA spectra, the oxygen migration in the Co grains is occurring both on the surface and throughout the bulk. In contrast, only the oxygen on the surface of the $GdO_x$ grains is likely mobile with the bulk of the grain remains stoichiometry balanced.

**Conclusion**

In summary, we report the first direct depth profile mapping of voltage-moderated oxygen migration in magnetic Co thin-films. Using X-ray and polarized neutron reflectometry we observed changes in the structural and magnetic profiles, consistent with partial oxidation of the Co layer. The oxidation and corresponding magnetic changes associated with application of a +40 V treatment were found to be strongest at the $GdO_x$/Co interface. We showed that the interfacial oxidation was largely reversible, and could be driven back to the Co layer with a reversed electric field, while the oxygen deeper in the film remained trapped. The effects of the electric field and thermal treatments were separated by comparing samples conditioned with and without an electric field. Magnetometry using the FORC technique revealed that the treatment altered the magnetic properties of the film, resulting in two distinct magnetic phases. X-ray spectroscopy revealed increased oxidation in the Co film after any conditioning, but less after the electro-thermal treatment with the reversed polarity. Thermal and electric cycling dramatically change the granular structure of the system, providing a means by which the $GdO_x$ can easily transport the oxygen. These results provide a depth-resolved view of magneto-ionic motion beyond the conventional interface limit, opening a new avenue to explore their applications in future device concepts.

**Methods**

Thin-film samples with a structure of Pd(40 nm)/$AlO_x$(1 μm) were fabricated by e-beam evaporation on naturally oxidized Si substrates in a high vacuum ($P_{Base}$ = $10^{-4}$ Pa) chamber. Deposition was performed at 150° C from nominally $Al_2O_3$ beads and 99.9% Pd ingots. Next, a 2 nm $GdO_x$ film was deposited by reactive DC sputtering (in another chamber with $P_{Base}$ = $10^{-4}$ Pa) in a 0.7 Pa $O_2$:Ar (1:3) atmosphere. During deposition of the $AlO_x$ and $GdO_x$, an edge was covered to allow for electrical contact to the Pd layer. Lastly, using a contact shadow mask, 1

cm$^2$ films of Co(15 nm)/Pd cap(20 nm) were deposited by DC sputtering in a 0.5 Pa Ar atmosphere. Using a shadow mask gives the film an oxide window-frame which prevents edge-shorting between the top and bottom Pd contacts. Electrical connections were made to the top and bottom Pd films by silver paint; the resistance through the thickness of the sample was measured to be >10 MΩ indicating minimal contributions from pinholes, if any.

The sample was characterized in the as-grown state, then electro-thermally conditioned by heating it to 230° C and applying a voltage of +40 V (400 kV/cm) - with the anode on the top surface - for 15 min. After measuring the sample in the conditioned state the sample was again heated to 230° C with a voltage of -40 V applied for 15 min. Control measurements were performed on identical films (grown side-by-side) where only heating was performed, hereafter referred to as thermally treated (or "thermal-only").

Polarized neutron reflectometry (PNR) measurements[41] were performed at the NIST Center for Neutron Research (NCNR) on the MAGIK and PBR reflectometers at room temperature with an in-plane applied magnetic field of 17 mT. Incident (scattered) neutrons were polarized with their spin parallel (+) or antiparallel (-) to the field, and the non-spin-flip specular reflectivities ($R^{\text{Incident Scattered}}$: $R^{++}$ and $R^{--}$) were measured as a function of wave vector transfer q. Magnetometry results measured at room temperature on a vibrating sample magnetometer show that the applied 17 mT field is large enough to saturate the cobalt magnetization. Thus the spin-flip reflectivities ($R^{+-}$ and $R^{-+}$) are zero, and are not considered here. The spin asymmetry representation of the data is calculated as SA = ($R^{++}$- $R^{--}$)/($R^{++}$+ $R^{--}$). Within the Born Approximation, the numerator of the SA is proportional to the product of the magnetization and nuclear SLD; for non-magnetic samples $R^{++}$ = $R^{--}$, thus SA = 0. Though it is a non-trivial mixture of magnetic and structural scattering, the SA can give a qualitative sense of the sample's magnetism. Model fitting of $R^{++}(q)$ and $R^{--}(q)$ allows for determination of depth (z) dependent nuclear and magnetic scattering length density (SLD).[42, 43] Model fitting of the PNR data was performed using the Refl1D software package.[44] In the model the Si, bottom Pd seed, and $AlO_x$ are slabs with a uniform SLD and no magnetic contribution. The $GdO_x$, Co and top Pd cap are modeled as a sum of error-function interfaces with fitted locations, widths and heights. A second model for the uncoated window frame was included as an incoherent sum. The data for the as-grown and conditioned states were simultaneously fitted with identical parameters for the Si, Pd seed and $AlO_x$ shared between the models. The nuclear SLD, $\rho_N$, presented in the main text are the mean-values, but appear in the fitted profile as a weighted average with neighboring layers at the interface. The neutron is sensitive to the net scattering potential at a defined depth, thus surface roughness and conformal roughness act to average the SLD at the interfaces.

FORC measurements were performed following previously outlined measurement procedures.[23, 24, 45-47] From positive saturation the magnetic field is decreased to a scheduled reversal field, $H_R$, then the magnetization, $M$, is measured as the applied field, $H$, is increased from $H_R$ back to positive saturation. This measurement process is repeated for a range of $H_R$ between positive and negative saturation, collecting a family of FORCs which fill the interior of the major loop. A mixed second order derivative is applied to extract the FORC distribution: $\rho(H, H_R) = -\frac{1}{2}\frac{\partial}{\partial H_R}(\frac{\partial M}{\partial H})$. To capture the reversible magnetic behavior, a constant extension is applied to the $H = H_R$ boundary of the dataset: $M(H<H_R) = M(H_R)$.[26] Recognizing that progressing to more negative values of $H_R$ probes down-switching events, and increasing $H$ probes up-switching, a new coordinate system is defined in terms of a local coercivity and bias field: $H_C = \frac{H-H_R}{2}, H_B = \frac{H+H_R}{2}$, respectively.

XA and XMCD measurements were performed at the Advanced Light Source on beamline 4.0.2. Elemental sensitivity was achieved by probing the Co $L_{2,3}$ edges, and Gd $M_{4,5}$ edge, following previously outlined procedures.[48-50] Measurements were performed using a constant beam polarization and an alternating in-plane magnetic field of ±200 mT. Signal was detected by fluorescence yield.

**Acknowledgements**

We would like to thank Prof. Geoffrey Beach from the Massachusetts Institute of Technology for illuminating discussions. D.A.G. and A.J.G. acknowledge support from the National Research Council Research Associateship Program. Work at UCD has been supported by the National Science Foundation (DMR-1543582 and ECCS-1232275). Work at the ALS has been supported by the Director, Office of Science, Office of Basic Energy Sciences of the U.S. Department of Energy (DEAC02-05CH11231).

**Figure Captions**

**Figure 1. PNR results of electro-thermally conditioned sample** (a) Fitted PNR data scaled by $q^4$ and (b) spin asymmetry for the sample as-grown and after electro-thermal conditioning. (c) Depth dependent nuclear and magnetic SLD, $\rho_N$ and $\rho_M$, extracted from the PNR. In (a) and (b) the experimental data are shown as symbols, and the lines are fits corresponding to the depth profile shown in (c). Error bars correspond to ±1 standard deviation.

**Figure 2. PNR results of thermally conditioned control sample.** (a) Fitted PNR profile scaled by $q^4$ and (b) spin asymmetry for the sample as-grown and after thermal-only conditioning. (c) Depth-dependent nuclear and magnetic SLD, $\rho_N$ and $\rho_M$, extracted from the PNR. In (a) and (b) the experimental data are shown as symbols, and the lines are fits corresponding to the depth profile shown in (c). Error bars correspond to ±1 standard deviation.

**Figure 3. Magnetometry and FORC investigations.** (a) Major hysteresis loops and FORC distributions for the sample (b) as-grown, (c) after +/- 40 V conditioning and (d) after thermal-only conditioning. The family of FORCs is shown in the insets.

**Figure 4. X-ray absorption and XMCD spectra.** XA (blue) and XMCD (red) spectra for (a) cobalt and (b) gadolinium in samples as-grown, after +/-40 V conditioning (E+T) and thermal-only conditioning. Arrows indicate 779.2 eV and 776.8 eV.

**Figure 5. Illustration of oxygen migration mechanism.** (Top) Cross-section view of grains during (left to right) +40 V and -40 V treatments, and after +/- 40 V treatment. Illustration

emphasizes fast surface migration and slow bulk migration. Bottom images show a cross section of the film, with migrating interstitial oxygen.

**Figures**

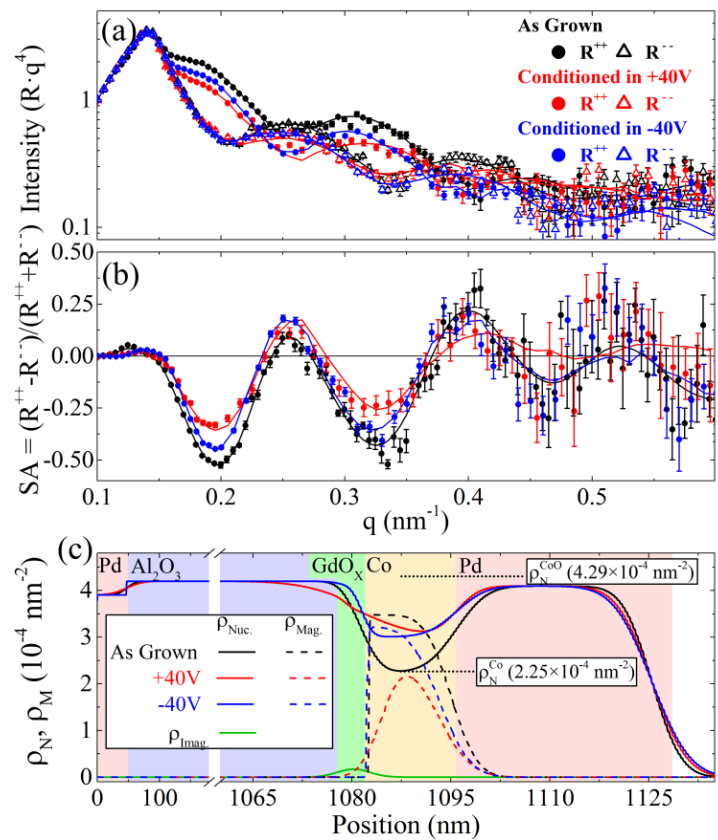

**Figure 1 Gilbert *et al.***

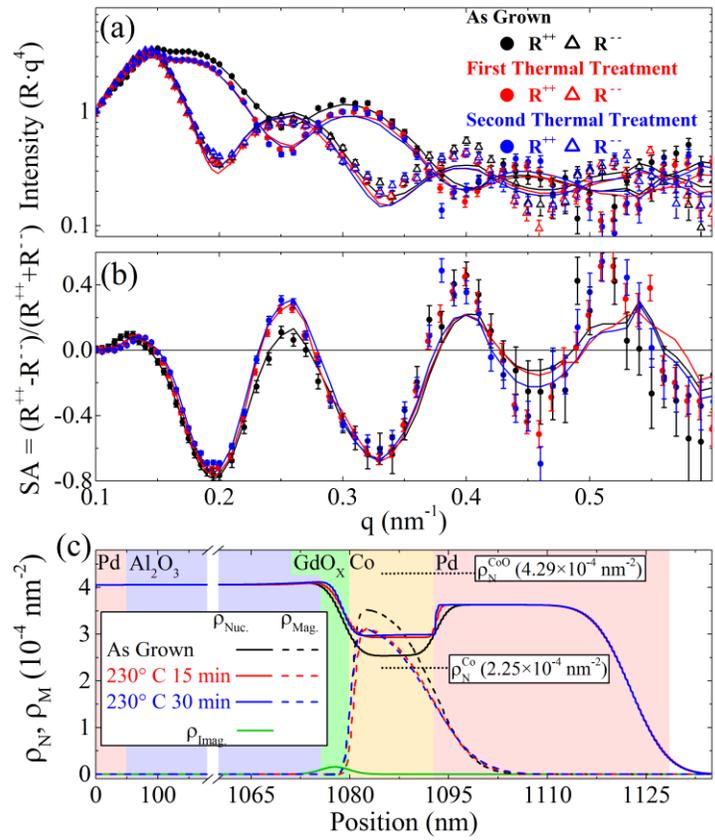

**Figure 2 Gilbert** *et al.*

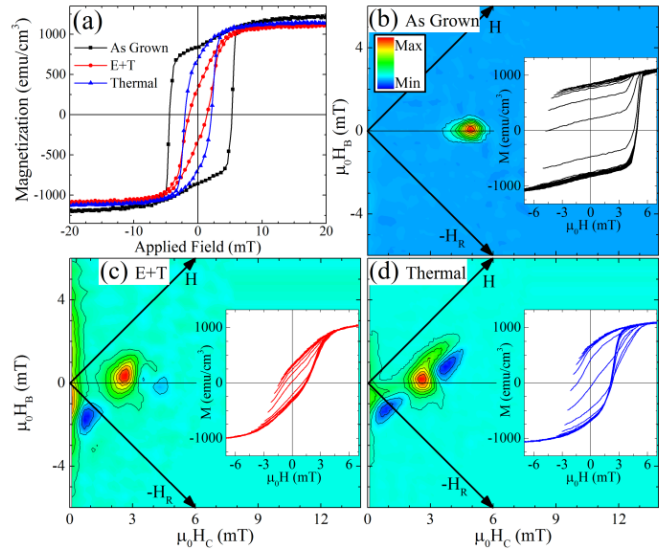

**Figure 3** Gilbert *et al.*

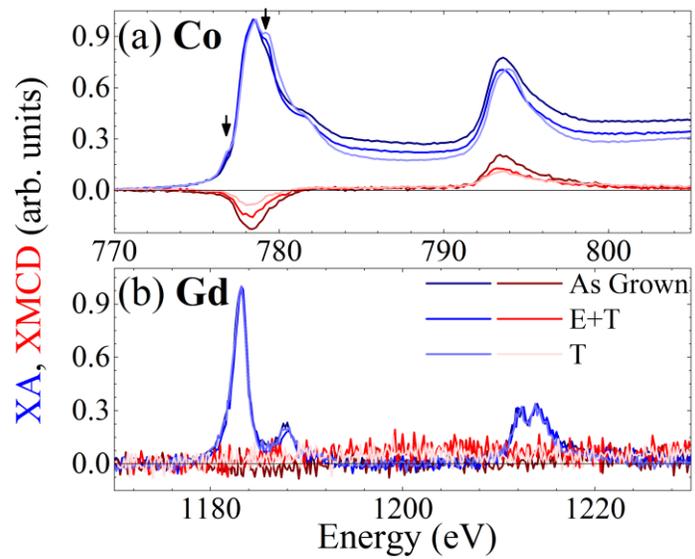

**Figure 4** Gilbert *et al.*

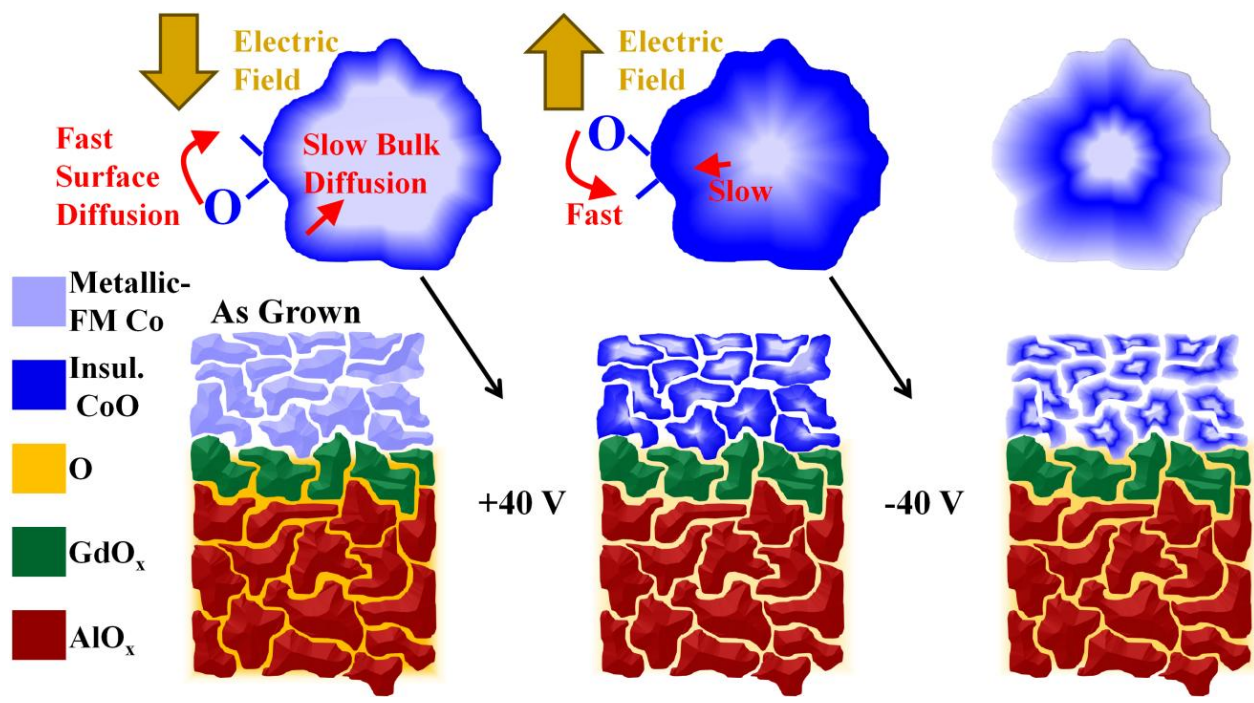

**Figure 5** Gilbert *et al.*

# Beyond the Interface Limit: Magnetic and Structural Profiles of Magneto-Ionic Controlled Heterostructures

Dustin A. Gilbert,[1] Alexander J. Grutter,[1] Elke Arenholz,[2] Kai Liu,[3] B. J. Kirby,[1] Julie A. Borchers,[1] Brian B. Maranville[1]

[1]NIST Center for Neutron Research, Gaithersburg, MD 20899

[2]Advanced Light Source, Lawrence Berkeley National Laboratory, Berkeley, CA 94720, USA

[3]Physics Department, University of California, Davis, CA 95616, USA

**Other Fitted Models**

A universal issue with fitting neutron data is the uniqueness of the fit. That is, for a given data set there may be several profiles which reproduce the experimentally measured scattering pattern. We have a high-level of confidence in our fits because (1) they reproduce approximately the accepted bulk values of nuclear and magnetic SLD for the constituent elements, (2) the measured structure agrees well with the designed structure, and (3) for a single sample the fits following different treatments were performed in parallel. Confidence factor (3) means that a model which is not physically accurate would not only have to reproduce one scattering pattern, but all three.

To further demonstrate confidence in the presented model, we attempted to fit the data with models with fixed magnetic and nuclear profiles. That is, the nuclear (magnetic) profile is constrained to be constant throughout all of the models, probing whether these results can be the result of only magnetic (nuclear) changes. In both cases the models would not converge to any reasonable value (e.g. physically unreasonable nuclear SLDs and interface widths and thicknesses significantly different from the designed structure).

**Other Possible Energies of Formation**

Below is a list of potential enthalpy of reactions for bulk gadolinium, cobalt and oxygen, with the calculated enthalpy of reaction.[32, 33] Positive enthalpy of reaction indicate the reaction is endothermic and thus must consume at least that much energy from the system. Negative enthalpy of reaction indicates exothermic. In order to be consistent with the observed results - which saw reversibility in the interfacial cobalt oxidation, by reversing the electric field - the enthalpy must be on a similar scale to the thermal + electric potential energy (67 meV).

$$GdO + Co \rightarrow CoO + Gd, \Delta H = -2.3 \text{ eV}$$

$$Gd_2O_3 + 3Co \rightarrow 3CoO + 2Gd, \Delta H = +8.6 \text{ eV}$$

$$Gd_2O_3 + Co \rightarrow CoO + 2GdO, \Delta H = +14.7 \text{ eV}$$

$$3GdO + 2Co \rightarrow Co_2O_3 + 3Gd, \Delta H = -5.26 \text{ eV}$$

$$Gd_2O_3 + 2Co \rightarrow Co_2O_3 + 2Gd, \Delta H = +12.9 \text{ eV}$$

$$4GdO + 3Co \rightarrow Co_3O_4 + 4Gd, \Delta H = -6.37 \text{ eV}$$

$$4Gd_2O_3 + 9Co \rightarrow 3Co_3O_4 + 8Gd, \Delta H = +47 \text{ eV}$$

All the above reactions - in addition to the ones presented in the paper - suggest that the chemical energies are much larger than the thermal and electric potential energy.

**Out-of-Plane Magnetometry**

Out-of-plane major hysteresis loops for the samples as-grown, after sequential +/-40 V and thermal-only treatments are shown in Fig. S1. The plot shows no significant hysteresis, and exceedingly small remanence and coercivity for all samples. This behavior identifies that the out-of-plane direction is the magnetic hard axis, as expected from the shape anisotropy. The largest hysteresis is observed in the as-grown sample, and may be the result of either residual magnetocrystalline anisotropy, or sample misalignment during measurement.

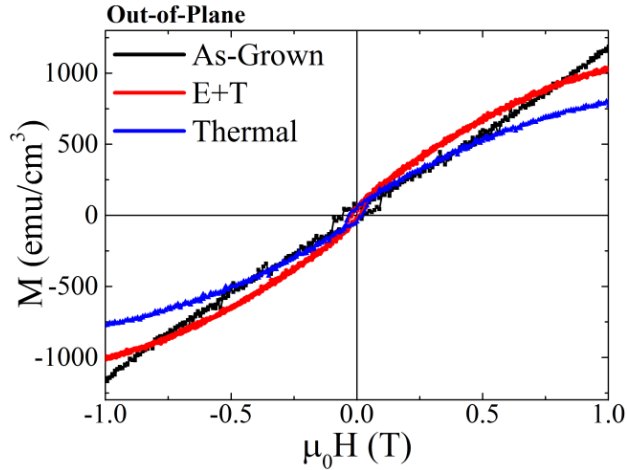

**Figure S1.** Out-of-plane hysteresis loops for the samples as-grown (black), after sequential +/-40 V (E+T, Red) and thermal-only treatments (Blue)

**Depth-Resolved Oxygen Profile**

In the main text, the oxygen content was calculated from the fitted SLD of the Co layer and the nuclear scattering lengths of Co and O. Two assumptions were made: (1) the as-grown Co-film was pristine, and (2) the number of Co atoms in the layer volume remained constant. Both of these are expected to be accurate based on the fitted SLD of the as-grown film and the fitted thicknesses. Expanding on (2) and assuming that the Co atoms do not migrate within the layer, the same approach can be used to calculate the depth-resolved oxygen profile, shown in Fig. S2. This plot shows explicitly the separate effects at the bottom 10 nm of the Co film, near the $GdO_x$/Co interface, and at the top 5 nm, near the Co/Pd interface. The thermal-only samples confirm oxidation after the first treatment and little additional oxidation after the second. This plot also shows the large difference between the electro-thermally treated samples and the thermal-only sample.

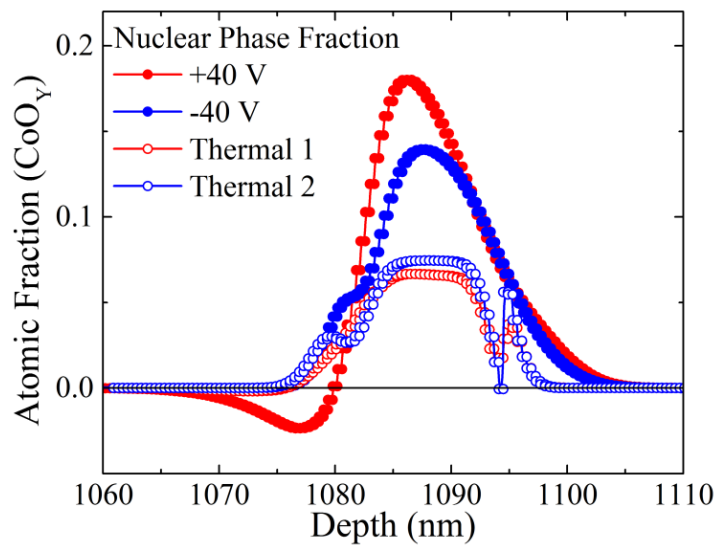

**Figure S2.** Calculated depth-resolved oxygen stoichiometry for the (solid) electro-thermally and (open) thermal-only conditioned samples, after their first (red) and second (blue) treatments.